# A Micro-PMU Placement Scheme for Distribution Systems Considering Practical Constraints

Reetam Sen Biswas, *Student Member, IEEE*, Behrouz Azimian, *Student Member, IEEE*, and Anamitra Pal, *Member, IEEE*.

*Abstract*— This paper presents an innovative approach to micro-phasor measurement unit (micro-PMU or μPMU) placement in unbalanced distribution networks. The methodology accounts for the presence of single-and-two-phase laterals and acknowledges the fact that observing one phase in a distribution circuit does not translate to observing the other phases. Other practical constraints such as presence of distributed loads, unknown regulator/transformer tap ratios, zero-injection phases (ZIPs), modern smart meters, and multiple switch configurations are also incorporated. The proposed μPMU placement problem is solved using integer linear programming (ILP), guaranteeing optimality of results. The uniqueness of the developed algorithm is that it not only minimizes the μPMU installations, but also identifies the minimum number of phases that must be monitored by them.

## I. INTRODUCTION

WITH the continuous addition of distributed energy resources (DERs) and active controllers occurring at the distribution levels, the power flowing through the distribution feeders is becoming increasingly uncertain [1]. Instances of unstable power supply, unintentional islanding, and voltage stability issues are manifesting more frequently [2]. Hence, there is a pressing need for real-time synchronized monitoring of distribution networks [3]. This has led to the creation of high precision micro-PMUs (μPMUs) [4] as well as modern smart meters [5] that can produce time-synchronized measurements.

Robust sensor placement methods for distribution networks proposed in recent literature have focused on frequent network reconfigurations [6] and relay operations [7]. Ref. [8] proposed an optimal placement scheme of μPMUs and conventional smart meters to ensure observability during contingencies. Optimal μPMU placement schemes for effective anomaly detection was investigated in [9]. In [10], the authors created an optimal measurement infrastructure using different devices for the distribution grid. In [11], μPMUs and intelligent electronic devices (IEDs) were optimally allocated for the distribution system using heuristic techniques. In [12]-[15], the μPMU placement problem was treated in a similar way as the optimal PMU placement problem for transmission systems.

The following attributes of the distribution network make optimal μPMU placement a more challenging problem than the optimal PMU placement problem for the transmission system: (i) mixture of single-phase, two-phase, and three-phase laterals, (ii) presence of distributed loads along the length of the feeders, (iii) zero-injection phases (ZIPs), (iv) unknown voltage regulator/transformer tap ratios, and (v) frequent changes in switch configurations. Prior literature [6]-[15] has not considered all the above-mentioned practical constraints simultaneously in their problem formulation.

Another limitation of prior research is the lack of distinction between node observability and phase observability. Unlike the transmission network, distribution networks have three-phase, two-phase, or single-phase nodes/feeders. Consequently, monitoring the distribution network translates to observing the phase voltages, and not the node voltages. Furthermore, prior research on μPMU placement only provided the node locations where the μPMUs were to be installed. Locating $n$ nodes does not necessarily imply that $n$ μPMUs would be required, as at a given node, more than one μPMU might be needed. The exact number of μPMUs to be installed at a node depends on the number of outgoing phases that must be monitored from that node location and the number of measurement channels that the μPMU has. For the distribution system, any sensor placement scheme is incomplete if it does not provide this vital information. That is, an optimal μPMU placement algorithm for the distribution system must minimize the combination of the number of μPMUs, node locations, and the number of phases that must be observed.

The primary objective of this research is to develop a μPMU placement algorithm subject to the above-mentioned practical constraints of the distribution system. The proposed algorithm is also generic enough to account for the presence of pre-installed unbundled smart meters (USMs) in the system. USMs are modern smart meters that can report time-synchronized data at the rate of one sample per second [5]. The relatively fast output rate of USMs compared to conventional smart meters as well as the time-synchronized nature of their measurements makes USMs suitable candidates for integration with μPMUs.

## II. THE NEED FOR A NEW MICRO-PMU PLACEMENT SCHEME

Since the transmission system is usually balanced, observability of one phase translates to observing all the other phases. However, most distribution networks are unbalanced, and often have only one-or-two phases present at a node. Therefore, observing one phase in a distribution system does not necessarily translate to observing the other phases. Fig. 1 shows the phase-connectivity between four nodes $i$, $j$, $k$, and $l$ of a distribution system. If node $k$ is to be indirectly observed by a μPMU, node $j$ must have preference over node $l$, because

This research has been funded by the ARPA-E research grant DE-AR-0001001.

phase A of node $k$ cannot be observed from node $l$, but all the three phase voltages of node $k$ can be observed from node $j$.

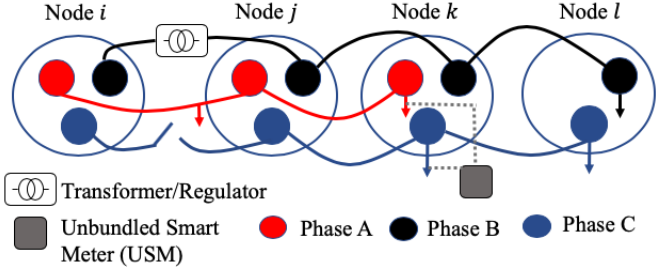

Fig. 1: Example of phase connectivity between different nodes of a distribution network.

Distributed loads are electrical loads that are present at different lengths along a feeder and are a unique characteristic of the distribution system. Fig. 1 shows the presence of distributed loads on phase A, along the feeder joining nodes $i$ and $j$. This means that if a μPMU is placed at node $i$, the A phase voltage of node $j$ cannot be observed from node $i$. As such, the presence of distributed loads along specific phases of a distribution feeder poses an additional constraint for distribution system observability. The same argument holds true for the presence of transformer/voltage regulators and switches on individual phases. Since the voltage regulator/transformer tap ratios vary frequently with system conditions, especially for a distribution system, the tap ratios should be treated as unknown. Therefore, placing a μPMU on one side of a voltage regulator or transformer does not translate to observing the other side. For example, the presence of a voltage regulator in the B phase of the feeder joining nodes $i$ and $j$, prevents a μPMU located at node $i$ to observe the B phase voltage at node $j$. Similarly, the distribution system is also characterized by the presence of "switches", which operate and change status at a much higher frequency than the transmission system. Since μPMU placement is a planning problem, an optimal μPMU placement scheme must ensure observability for all "feasible" switch configurations (see Section III for the definition of a feasible configuration). For example, if a μPMU is placed at node $i$, the presence of a "switch" in phase C of the feeder between nodes $i$ and $j$, hinders the observability of the C-phase voltage at node $j$ from node $i$, if the switch is open.

Concept of zero-injection buses (ZIBs) has been exploited in the transmission system for reducing the number of installed PMUs [16]. The concept of ZIBs gives way to the concept of zero-injection phases (ZIPs) for distribution networks. This is because at a given node among the three phases only a subset of the phases might have zero-injections. Considering that a μPMU is placed at node $j$, and phase B of node $k$ does not have any spot load (or injection), the phase B voltage of node $l$ can be observed from the μPMU at node $j$. However, if there is a "spot load" present, as in the phase C of node $k$, the C-phase voltage of node $l$ cannot be observed from node $j$.

Modern smart meters such as USMs also influence network observability. Consider now that node $k$ of Fig. 1 is indirectly monitored by a μPMU placed at node $j$, and a USM is placed at node $k$ that provides active and reactive power injection information of the spot loads at node $k$. With the knowledge of the three-phase voltage phasor at node $k$ (from the μPMU at node $j$), and the active and reactive power demand at node $k$ (from the USM at node $k$), the current phasor from node $k$ to node $l$ can be calculated. Consequently, the C-phase voltage at node $l$ can be found, without placing a μPMU at node $k$ or node $l$. In essence, having one or more phase voltages monitored by a USM has a similar effect on observability as a ZIP.

## III. MATHEMATICAL FORMULATION

Let the power distribution system be represented by an undirected graph $\mathcal{G}(V, E)$, where $V$ is the set of nodes and $E$ is the set of edges. The μPMU placement must be done with the objective that every phase at every node of the system is observed. To attain this objective, the original graph $\mathcal{G}$ must be modified to account for the individual phases. Hence, from the graph $\mathcal{G}$ we form the graph $\mathcal{G}'(V', E')$, where $V'$ represents the set of phases, and $E'$ represents every edge that joins a distinct phase of one node to the same phase of another node. Each element of the set $V'$ is represented by a pair of numbers $(x, y)$ such that $x \in \{1, 2, \ldots, M\}$, where $M = |V|$ and $y \in \{1, 2, 3\}$, where 1 refers to A-phase, 2 refers to B-phase, and 3 refers to C-phase inside node $x$. Next, a lexicographic ordering scheme is introduced among the phases. For the two phases $v_1 = (x_1, y_1)$ and $v_2 = (x_2, y_2)$, $v_1 \prec v_2$ if $x_1 < x_2$. Every edge $e \equiv \{v_1, v_2\} \in E'$ joins two phases $v_1 = (x_1, y_1)$ and $v_2 = (x_2, y_2)$ from two different nodes $x_1$ and $x_2$. In subsequent notations, if an edge $e \equiv \{v_1, v_2\}$ is specified, it will be assumed that $v_1 \prec v_2$, implying that $v_1$ is the low end and $v_2$ is the high end of the edge $e$.

Every node $x_i \in V$ for $1 \leq i \leq M$ is now associated with a binary variable $z_i$ such that

$$z_i = \begin{cases} 1 & \text{if μPMU is placed on node } x_i \\ 0 & \text{otherwise} \end{cases} \quad (1)$$

An integer variable $n_i$ is also associated with every node $x_i$, such that $n_i$ denotes the number of μPMUs placed inside node $x_i$. Now, for any phase $v \in V'$ present at a given node, let $L_v$ denote the set of edges for which $v$ is the "low end". Similarly, let $H_v$ be the set of edges for which $v$ is the "high end"; that is $E'_v = L_v \cup H_v$. Then, every edge $e$ can be associated with two binary valued variables $g^l_e$ and $g^h_e$ such that [16]

$$g^l_e = \begin{cases} 1 & \text{if a μPMU observes the low end of egde } e \\ 0 & \text{otherwise} \end{cases} \quad (2)$$

$$g^h_e = \begin{cases} 1 & \text{if a μPMU observes the high end of edge } e \\ 0 & \text{otherwise} \end{cases} \quad (3)$$

Based on (1)-(3), the objective function for μPMU placement would be expressed as shown below.

$$\text{Minimize} \left( \sum_{i=1}^{M} z_i + \sum_{e \in E'} (g^l_e + g^h_e) + \sum_{i=1}^{M} n_i \right) \quad (4)$$

Equation (4) simultaneously minimizes the number of affected nodes, the number of monitored phases, and the total number of μPMUs. The optimization problem is formulated as an integer linear programming (ILP) problem that guarantees globally optimum solutions. The different constraints applied to the optimization problem are described below.

1. *Phase observability constraint (binary variables $g_e^l$, $g_e^h$):*
   For every phase $v \in V'$, the constraint for phase observability is shown below [16].

$$\sum_{e \in E_v'} (g_e^l + g_e^h) \geq 1 \quad (5)$$

2. *Constraint for number of µPMUs (integer variable $n_i$):*
   Let a µPMU have $K$ measurement channels, where one channel can measure one voltage and one current phasor. Also, let $V_i' \subseteq V'$ contain the phases that are present at node $x_i$. Then, number of µPMUs, $n_i$, to be placed at node $x_i$ is given by

$$n_i \geq \frac{\sum_{v \in V_i'} \sum_{e \in L_v} g_e^l + \sum_{v \in V_i'} \sum_{e \in H_v} g_e^h}{K} \quad (6)$$

3. *Constraint for the affected nodes (binary variable $z_i$):*
   For every phase $v \in V'$, and every edge $e \in L_v$ or $e \in H_v$, following constraints must be added for $z_i$ (which corresponds to the node $x_i$) [16].

$$\left. \begin{array}{l} z_i \geq g_e^l, \quad \forall e \in L_v \\ z_i \geq g_e^h, \quad \forall e \in H_v \end{array} \right\} \quad (7)$$

It is important to highlight here that the binary variable $z_i$ gives the node locations where the µPMUs must be installed, the variables $g_e^l$ and $g_e^h$ give the outgoing phase laterals that must be monitored from a given node, and the integer variable $n_i$ gives the number of µPMUs contained at the node $x_i$.

The additional constraints (introduced in Section II) are modeled as follows:

1. *Handling of voltage regulators/transformers and distributed loads:* It has been shown in Section II that the presence of unknown voltage regulator/transformer tap ratios and distributed loads can influence network observability. Therefore, such additional constraints must be accounted for during the µPMU placement formulation itself. If the edge set $E_{VR} \subseteq E'$ includes the voltage regulators or transformers, then, all the edges $e \in E_{VR}$ must be removed from the edge set $E'$ of the modified graph $\mathcal{G}'$ before (1)-(7) are implemented. Similarly, if distributed loads are present on the set of edges $E_{DL}$, where $E_{DL} \subseteq E'$, it implies that all edges $e \in E_{DL}$ must be removed from the edge set $E'$ of the modified graph $\mathcal{G}'$ before the optimization problem is solved.

2. *Handling of different switch configurations:* Here, the objective is to ensure complete phase observability for all feasible switch configurations. If there are $s$ switches, $2^s$ switch configurations are possible. However, in the context of this research, only those switch configurations are deemed feasible, which do not result in an islanded mode of operation. For a given system, let there be $f$ such feasible switch configurations, implying that there exist $f$ connected graph topologies, $\mathcal{G}_1', \mathcal{G}_2', \ldots, \mathcal{G}_f'$. The constraint equations (5)-(7) for a given topology can be grouped together and written in the form shown below.

$$AX \geq B \quad (9)$$

where $A$ contains the coefficients of the different integer variables ($z_i$, $g_e^l$, $g_e^h$, and $n_i$), $X$ contains all the integer variables, and $B$ contains the constants of each constraint equation. If for each of the $f$ network topologies, $\mathcal{G}_1', \mathcal{G}_2', \ldots, \mathcal{G}_f'$, the respective $A$ and $B$ matrices are $A_1, A_2, \ldots, A_f$, and $B_1, B_2, \ldots, B_f$, the µPMU placement solution will satisfy all $f$ switch configurations if the following holds true.

$$\begin{bmatrix} A_1 \\ A_2 \\ \vdots \\ A_f \end{bmatrix} X \geq \begin{bmatrix} B_1 \\ B_2 \\ \vdots \\ B_f \end{bmatrix} \quad (10)$$

3. *Handling of ZIPs and USMs:* The concept of zero-injections is handled in a manner similar to what was done in [16]. Let $V_{ZI}$ denote the set of all phases whose injections are known (either by them being a ZIP or through a USM). Let the neighborhood of a phase $v \in V'$, denoted by $N_v$, contain the phase $v$ itself and all phases that are adjacent to $v$. Now, we define $V_{ZIS} \subseteq V_{ZI}$, such that, $V_{ZIS}$ contains only those known injection phases that are at the same voltage level as other nodes in their neighborhood, and do not contain distributed loads between themselves and any other phase in their neighborhood. Let $m = |V_{ZIS}|$. An object set $R$ defined for modeling the observability constraints, is described below:
For $m = 1$: If $i$ is a single element of the set $V_{ZIS}$ and $N_i$ represents the neighborhood of phase $i$, for every pair of elements $p, q \in N_i$, the object set $R$ contains the two-element set $\{p, q\}$.
For $m \geq 2$: Let $j$ and $k$ be any two elements of set $V_{ZIS}$, such that sets $N_j$ and $N_k$ denote the neighborhoods of $j$ and $k$, respectively. It is important to note that the sets $N_j$ and $N_k$ may have elements in common. Let $N_{j,k}$ represent the common elements of the sets $N_j$ and $N_k$; i.e., $N_{j,k} = N_j \cap N_k$. Let $N_j'$ contain the elements that are present only in set $N_j$, but not in set $N_k$; i.e., $N_j' = N_j - N_{j,k}$. Similarly, let $N_k'$ contain the elements that are present only in set $N_k$, but not in the set $N_j$; i.e., $N_k' = N_k - N_{j,k}$. For each pair of elements $p$ and $q$ in $N_j'$ or $N_k'$ or $N_{j,k}$, the two-element set $\{p, q\}$ is added to $R_{j,k}$. Next, the cross-product set $Q_{j,k} = N_j' \times N_k' \times N_{j,k}$ is calculated, which consists of all the triplets $(p, q, r)$ such that $p \in N_j'$, $q \in N_k'$ and $r \in N_{j,k}$. Each triplet $(p, q, r)$ is added to the object set $R_{j,k}$. For each pair of elements $j$ and $k$, the object set $R_{j,k}$ is created for every pair. The total collection of objects stored in the object set $R$ is given below [16].

$$R = \bigcup_{1 \leq j \leq k \leq m} R_{j,k} \quad (11)$$

In (11), the $\cup$ operator eliminates duplicate entries. For every set $(p, q) \in R$ and $(p, q, r) \in R$ the modified observability constraints are given below [16].

$$\left. \begin{array}{l} \sum_{e \in E_p'} \{g_e^l + g_e^h\} + \sum_{e \in E_q'} \{g_e^l + g_e^h\} \geq 1 \\ \sum_{e \in E_p'} \{g_e^l + g_e^h\} + \sum_{e \in E_q'} \{g_e^l + g_e^h\} + \sum_{e \in E_r'} \{g_e^l + g_e^h\} \geq 1 \end{array} \right\} \quad (12)$$

## IV. RESULTS AND DISCUSSION

The ILP problem was solved in MATLAB using Gurobi as the optimizer. For the simulations done in this paper, we have assumed that (i) a subset of the phases that have relatively large spot loads are monitored by USMs, and (ii) a μPMU has three measurement channels and can therefore measure three voltage and three current phasors [17]. The proposed μPMU placement algorithm not only finds the node locations where the μPMUs must be installed, but also the phases that must be monitored. This is explained in Table I using the IEEE 13-node distribution feeder shown in Fig. 2. In this system, without considering ZIPs, seven μPMUs are placed at five node locations, namely, 2, 4, 7, 9, and 12 with the consideration that USMs are located at nodes 5, 6, 8, and 10.

The μPMU placement results for four IEEE distribution feeders are summarized in Table II. The second column indicates whether ZIPs were considered in the analysis or not. The third column indicates the total number of ZIPs in comparison to the total number of phases present in the system. In IEEE 13-node distribution feeder there exists 18 ZIPs and 32 phases. The fourth and the fifth columns give the number of node locations required for μPMU installation and the number of μPMU devices, respectively. The last column of Table II gives the locations of USMs present in the systems. It is observed that for the IEEE 13, 34, 37, and 123 node systems, on considering ZIPs, the number of μPMUs reduced from 7 to 6, 25 to 22, 19 to 15, and 54 to 43, respectively.

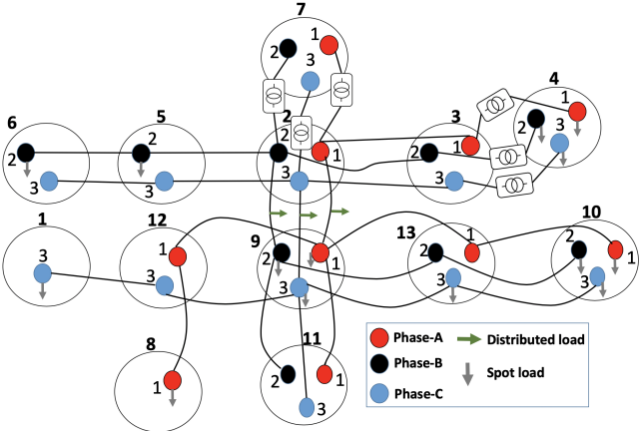

Fig. 2: Phase-connectivity between different nodes of the IEEE 13 node distribution feeder (nodes renumbered in ascending order).

Tables III and Table IV provide the μPMU locations for the test systems, without considering ZIPs and with considering ZIPs, respectively. The way in which the consideration of ZIPs reduces the number of μPMUs is explained below, with regards to the IEEE 13-node distribution feeder.

- The μPMUs that are positioned at nodes 2, 4, and 7 monitor the same phases that were monitored when ZIPs were not considered (compare Table I with Table V). The presence of a transformer and a voltage regulator between nodes 2,7, and 3,4, respectively, necessitates the installation of μPMUs at each of the nodes 4 and 7.
- The A, B, and C phases of node 11 do not have any load. Therefore, there is zero incoming current at node 11 from node 9, implying that the three-phase voltage of node 11 would be the same as that of node 9, because of zero voltage drop in the feeder 9-11. Consequently, there is no need to monitor the phase laterals (9,1)-(11,1), (9,2)-(11,2) and (9,3)-(11,3), which were monitored when ZIPs were not considered (see Table I).
- The presence of a USM at node 5 allows the phase voltages of node 6 to be observed by a μPMU placed at node 2. Similarly, the presence of a USM at node 10 allows the μPMU placed at node 9 to observe node 10.
- The A and C phases of node 12 do not have any net injection. One μPMU placed at node 9 indirectly monitors the A and C phases of node 12. Therefore, we do not need a separate μPMU for observing the edge (12,3)-(1,3) as the current that flows from (12,3) towards (1,3), must be the same as the current that flows from (9,3) to (12,3) (refer Fig. 2). Since, edge (9,3)-(12,3) is monitored by a μPMU, C phase voltage of node 1 can be effectively observed. The same reasoning applies for edges (9,1)-(12,1) and (12,1)-(8,1), because the A phase of node 12; i.e., (12,1) is a ZIP.

Table I: μPMU installation for IEEE 13-node feeder (without ZIP)

| Node Location | Phases Monitored** | # μPMU | USM location |
|---|---|---|---|
| 2 | (2,1)-(3,1); (2,2)-(3,2); (2,3)-(3,3); (2,2)-(5,2); (2,3)-(5,3) | 2 | 5,6,8,10 |
| 4 | (4,1)-(3,1); (4,2)-(3,2); (4,3)-(3,3) | 1 | |
| 7 | (7,1)-(2,1);(7,2)-(2,2);(7,3)-(2,3) | 1 | |
| 9 | (9,1)-(11,1);(9,2)-(11,2);(9,3)-(11,3); (9,1)-(13,1);(9,2)-(13,2);(9,3)-(13,3) | 2 | |
| 12 | (12,1)-(8,1);(12,3)-(1,3) | 1 | |

** The first and second term inside the bracket denote the node number and phase numbers (1 for Phase-A, 2 for Phase-B, 3 for Phase-C), respectively.

Table II: μPMU placement results for IEEE distribution feeders

| Test System | ZIP+ | #ZIP (#Phases) | # Nodes | # μPMU | USM Locations |
|---|---|---|---|---|---|
| 13-node | No | 18 (32) | 5 | 7 | 5,6,8,10 |
| | Yes | | 4 | 6 | |
| 34-node | No | 68 (86) | 22 | 25 | 22 |
| | Yes | | 21 | 22 | |
| 37-node | No | 79 (111) | 14 | 19 | 1,15,17,24,26,30, 31,34 |
| | Yes | | 13 | 15 | |
| 123-node | No | 176 (273) | 51 | 54 | 22,43,47,48,64,7 7,80,90,106,107 |
| | Yes | | 41 | 43 | |

Table III: Optimal node locations (not considering ZIP)

| Test System | Node Locations where μPMUs are placed |
|---|---|
| 13-node | 2,4,7,9,12 |
| 34-node | 1,4,5,6,8,10,11,13,14,16,18,19,20,21,23,24,27,28,29,30, 32,33 |
| 37-node | 2,3,5,7,8,9,10,11,14,19,30,35,36,37 |
| 123-node | 1,3,6,8,13,14,15,18,19,23,27,28,29,32,36,39,40,44,46,49, 51,52,54,55,58,60,63,65,67,70,74,76,78,82,84,87,91,95, 97,99,101,103,105,110,113,115,117,123,124,127,128 |

Table IV: Optimal node locations (considering ZIP)

| Test System | Node Locations where μPMUs are placed |
|---|---|
| 13-node | 2, 4,7, 9 |
| 34-node | 1,4,5,8,10,11,13,14,16,17,19,20,23,24,27,28,29,30,31,32, 33 |
| 37-node | 2,3,5,6,7,8,10,11,14,27,35,36,37 |
| 123-node | 5,8,14,15,18,20,21,25,27,31,35,38,42,46,49,52,54,58,60, 63,65,67,70,75,76,78,82,85,87,91,95,97,100,103,105,109, 113,116,118,122,124 |

Table V: μPMU installation for IEEE 13-node feeder (considering ZIP)

| Node Location | Phases Monitored | # μPMU | USM Locations |
|---|---|---|---|
| 2 | (2,1)-(3,1); (2,2)-(3,2); (2,3)-(3,3) (2,2)-(5,2); (2,3)-(5,3) | 2 | 5, 6, 8,10 |
| 4 | (4,1)-(3,1);(4,2)-(3,2);(4,3)-(3,3) | 1 | |
| 7 | (7,1)-(2,1); (7,2)-(2,2); (7,3)-(2,3) | 1 | |
| 9 | (9,1)-(12,1); (9,3)-(12,3); (9,1)-(13,1);(9,2)-(13,2);(9,3)-(13,3) | 2 | |

The unique advantage of simultaneously minimizing the number of affected nodes and the total number of μPMUs is discussed next. Consider the 5-node system depicted in Fig. 3. The squares marked with letter "p" denote the position of the μPMU channels as obtained using the proposed methodology. It is observed that two nodes (nodes 2 and 3) were disrupted for μPMU installations, the number of monitored phase laterals were eight, and the number of μPMUs required is three (two at node 3 and one at node 2). The results that one will obtain if the two above-mentioned objectives are not minimized simultaneously are described below.

*Case A-Minimizing only the number of node locations*: If the number of μPMUs are not included inside the objective function, then the placement solution that will be obtained is denoted by the squares marked with letter "a" in Fig. 3. This placement scheme also affects two nodes (nodes 2 and 3) and monitors eight phase laterals. However, four μPMUs are required in this scenario: three at node 3 and one at node 2 (since a μPMU has three measurement channels).

*Case B-Minimizing only the number of μPMUs*: If the number of affected node locations are not included inside the objective function, then the placement solution that will be obtained is given by the squares marked with letter "b" in Fig. 3. This placement scheme uses three μPMUs to monitor eight phases; but it affects three nodes, as one μPMU is installed at each of the nodes 2, 3, and 4. Installing three μPMUs at two different node locations is better than installing three μPMUs at three separate node locations. This is because the associated infrastructure costs are proportional to the number of sites (nodes) that are disrupted for device placement [16].

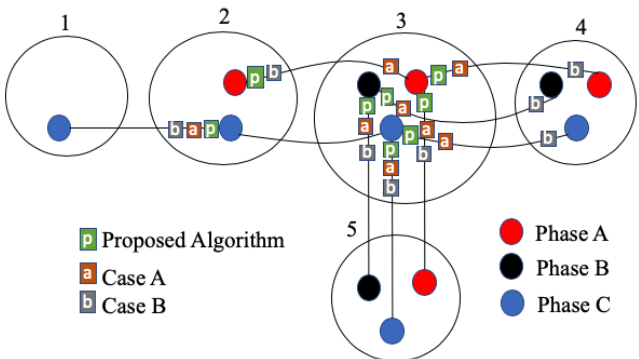

Fig. 3: μPMU placement solutions subject to different objectives.

## V. CONCLUSIONS AND FUTURE WORK

This paper proposed an optimal μPMU placement scheme that is subject to the practical constraints of a distribution system. The optimization problem was formulated as an ILP, which guaranteed optimality of results. The proposed optimization methodology can handle the presence of single- and-two-phase laterals, distributed loads, modern smart meters, unknown tap ratios of voltage regulators/transformers, different switch configurations, and ZIPs, simultaneously. Prior placement schemes only identified the node locations for μPMU installation. The proposed research not only finds the optimal node locations, but also the minimum number of μPMU devices required at a node, and the least number of phases that must be monitored by them. The importance of considering "the phases to be monitored" for distribution system state estimation will be described in a future publication. Multistage placement of μPMUs will also be investigated as a future topic of research.